\documentclass[pre,twocolumn,amsmath,amssymb,showpacs]{revtex4}
\def\dbar{{\mathchar'26\mkern-12mu d}}
\usepackage{graphicx}
\usepackage{dcolumn}
\usepackage{soul}
\usepackage{bm}

\begin{document}

%
\title{Local Nonequilibrium Configurational Entropy in Quasi-one-dimensional Heat Conduction}

\author{Gary P. Morriss$^{~\star}$}

\affiliation{School of Physics, University of New South Wales, Sydney NSW 2052, Australia}
$^{~\star}$ E-mail: (g.morriss@unsw.edu.au). \\[12pt]


\date{\today}

\begin{abstract}
In a quasi-one-dimensional system the particles remain ordered from left to right allowing the association of a volume element to the particle which on average resides there. Thus the properties of that single particle can give the local densities in the volume element. With reservoirs of different temperatures connected to each end of the system a steady heat current with an anomalous thermal conductivity results. A local configurational entropy density is calculated from  two-particle correlation functions which varies locally within the nonequilibrium steady state. This local configurational entropy is proposed as the  configurational component of the local entropy of the nonequilibrium steady state. 
\end{abstract}


\maketitle  

\section{Introduction}

   For a system of $N$ particles in volume $V$ at temperature $T$ in thermal and mechanical equilibrium the true entropy is the Gibbs entropy \cite{Green}
\begin{equation}\label{1}
S= - \frac{k} {N!} \int f_{\Gamma} (\Gamma) \ln f_{\Gamma}(\Gamma) d\Gamma= \frac {U} {T} + k \ln Z_N
\end{equation}
where $f_{\Gamma}(\Gamma)$ is the probability that an $N$ particle system has a particular set of particle positions and velocities $\Gamma =( r^{N},v^{N})$. 
   $U$ is the internal energy and $k$ is Boltzmann's constant.

   If the distribution $f_{\Gamma}(\Gamma)$ factors into the product of a velocity distribution $f_{N}^{(N)} (v^{N})$ (which is Gaussian at equilibrium) and a configurational $N$-particle distribution $n_{N}^{(N)} (r^{N})$, where the $(N)$ superscript signifies that the system contains $N$ particles, then at equilibrium it is usual to define a new configurational correlation function \cite{Stell} for a sub-set of $K$ particles $g_{K}^{(N)}$ as
\begin{equation}\label{gr}
g_{K}^{(N)} (1,....,K) = \frac {n_{K}^{(N)} (1,....,K)} {\prod n_{1}^{(N)} (i)}
\end{equation}   
(where $n_{1}^{(N)} (i)$ is the local number density at the position $i$). 
   The correlation function $g_{K}^{(N)}$ has the property that it approaches one as both the separation between particles, and the system size, becomes infinite.

   The first attempt to calculate the configurational entropy of a dense fluid at equilibrium was by Green \cite{Green}. Considering a canonical ensemble of $N$ particles, he argued that a hierarchy of correlation function ratios of the form
\begin{equation}\label{prod}
\delta g_{K}^{(N)} (1,....,K) = \frac {g_{K}^{(N)} (1,....,K)} {\prod g_{K-1}^{(N)} (\alpha_{1},....,\alpha_{K-1})}
\end{equation}   
are closer and closer to one as $K$ increases.
   The product is over all distinct choices of $K-1$ symbols from the set $\{1,...,K\}$. As there are $K$ choices of the label to leave out, there are $K$ terms in the product.
   Taking the logarithm of Eqn. (\ref{prod}) and solving recursively we obtain an expansion for the term $ \ln n^{(N)}_{N}$ implicit in Eq. (\ref{1}).

   Nettleton and M. S. Green \cite{NG58}, neglecting correlations between three or more particles, obtain a similar result for the configurational entropy in the grand canonical ensemble. Raveche \cite{R71} extended these results to four particle correlations discussing both open (grand canonical) and closed (canonical) systems. In the following paper Mountain and Raveche \cite{MR71} present the first modern calculations of the configurational entropy based on the PY approximation for hard spheres \cite{PY}. Implicit in many of these early works is the idea that the expressions obtained may be useful for calculating the entropy of nonequilibrium systems \cite{Green, NG58}. 
   Later, in a series of papers Wallace \cite{W87} applied these methods to the calculation of configurational entropy for  real systems such as liquid sodium.

   For an equilibrium system in the grand canonical ensemble the entropy per particle \cite{BB92} is
\begin{eqnarray} \label{greenent}
\frac {S} {Nk} &=& -\int f(v) \ln f(v) dv - \ln \rho  \nonumber \\
 &-& \frac {1} {2} \rho \int d{\bf r} (g_{2} \ln g_{2} - g_{2} + 1) \nonumber \\
 &-& \frac {1} {6} \rho^{2} \int \int d{\bf r} d{\bf r}' (g_{3} \ln \delta g_{3} -g_{3} + 3 g_{2} g_{2} - 3 g_{2} +1) \nonumber \\
 &-& .....
 \end{eqnarray}
where $\rho$ is the density which is uniform at equilibrium.
   The first two terms arise from kinetic theory. 
   There is a kinetic contribution from the velocity distribution and the term $-\ln \rho$ is the configuration term that appears in kinetic theory and the Sackur-Tetrode equation from the local equilibrium distribution.
   The term involving 2-body correlations we will refer to as $s^{\phi}$.
   The remaining term is the configurational contributions from 3-body correlations but higher order correlation function contributions also exist.
    Banayai and Evans \cite{BE91} have shown that when compared with thermodynamic integration, a numerical calculation at equilibrium using 2-body correlations typically yields approximately $90\%$ of the entropy and including 3-body correlations gives most of the rest.
    In all results presented here we consider only 2-body correlations, neglecting contributions from 3 (or more)-body  correlations.

   Molecular dynamics simulations have proved a very effective means of testing theoretical approaches to the study of fluids both in equilibrium, and nonequilibrium steady states \cite {EM2,D13}. 
   For hard particles, in the absence of external forces, a particle trajectory is linear between collisions, so numerical simulations are only limited in accuracy by wordlength, and the accuracy of averages is only limited by statistical considerations. 


   In kinetic-theory the basic ingredient is the Boltzmann entropy $S(t)$ which is defined, up to a constant, to be
\begin{eqnarray}\label{Bent}
S(t)&=&\int d{\bf r}~s({\bf r},t)   \nonumber \\
&=& - \int d{\bf r} \int d{\bf v}~ f_{\mu}({\bf r},{\bf v},t)\ln f_{\mu}({\bf r},{\bf v},t)
\end{eqnarray}
where $s({\bf r},t)$ is the entropy density at position ${\bf r}$ at time $t$. 
   The time evolution of the $\mu$-space distribution function $f_{\mu}({\bf r},{\bf v},t)$ can be obtained from the Boltzmann equation which takes the following form 
\begin{equation}\label{Beq}
\frac{\partial f_{\mu}}{\partial t}+ {\bf v}\cdot \frac{\partial f_{\mu}}{\partial {\bf r}}+ {\bf F}^{e}\cdot \frac{\partial f_{\mu}}{\partial {\bf v}}=J[f_{\mu}],
\end{equation}
where ${\bf F}^{e}$ is the external force as internal forces are included in the collision integral $J[f_{\mu}]$. 
   In the absence of external forces, and when the spatial distribution $f(r)$ is uniform, the distribution function $f_{\mu} \rightarrow \rho f({\bf r}))$ which is normalized as
\begin{equation}\label{dist}
\int d{\bf v}f({\bf r},{\bf v},t)=n({\bf r},t),
\end{equation}
where $n({\bf r},t)$ is the local number density of the system.
   The kinetic contributions to the entropy of the quasi-one-dimensional (QOD) system have been studied recently \cite{KM09} and one of the purposes of this paper is to show that the configurational contributions can also be estimated.

   The relation between the entropy flux and the heat flux may be viewed as a generalized version of the equilibrium Clausius relation $\dbar Q = TdS$. 
   It was found \cite{KM09} that the kinetic fluxes of heat and entropy times the local temperature match locally for the QOD system away from equilibrium and the kinetic entropy density agrees very well with the local equilibrium result.
   One of the goals of this work is to determine whether this matching extends to the configurational components.
   The concept of temperature and entropy away from equilibrium are active open problems \cite{Jou01,CVJ03,Jou11}. 
           

\section{The Model System}

\begin{figure}[htb]
\begin{center}
\caption{Schematic presentation of an $N$ hard-disk quasi-one-dimensional (QOD) system. The height $L_y$ is sufficiently small that the disks cannot pass one another. We choose the coordinate origin to be located at the bottom left corner of the system, and the periodic upper and lower system boundaries at $y=0,L_y$ are denoted by dashed lines. The boundaries at $x =0$ and $x=L_x$ are the hard walls of the reservoirs.} 	\label{QDS}
	\includegraphics{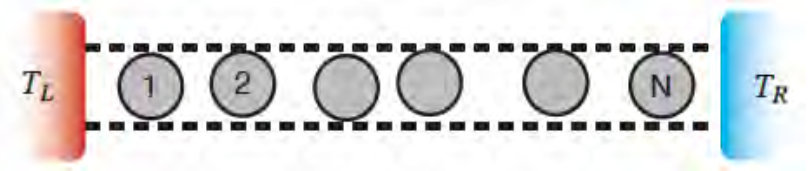}
\end{center}
	\end{figure}

      The quasi-one-dimensional (QOD) system of hard disks introduced in \cite{TM03} (see figure (\ref{QDS})) can be modified to interact with an idealized {\it heat  reservoir} in a deterministic and reversible way. 
   The deterministic thermal reservoir couples to the QOD system of hard disks by changing the collision rule at the reservoir boundary \cite{TM07}. 
   A reservoir collision preserves the tangential component of momentum but the normal $x$-component after collision becomes
\begin{equation} \label{wallcoll}
p'_{x} = \epsilon  p_{res}-(1- \epsilon)  p_{x},
\end{equation}	 
where $p_{res}$ is the fixed value of the reservoir momentum determined by the reservoir temperature $p_{res}=\sqrt{ 2T_{res}}$ and $\epsilon$ is a reservoir coupling parameter.
   As $\epsilon \rightarrow 0$ the system decouples from the reservoir and the boundary becomes a hard wall, and as $\epsilon \rightarrow 1$ the incoming momentum is replaced by the reservoir momentum.
   Recent studies of this system \cite{KM09,MT12,MT13} have shown that when in contact with two reservoirs of the same temperature the active mechanical coupling leads to a kinetic entropy production near each reservoir boundary which then flows into the reservoir.
   This effect is local and restricted to a small number of boundary layer particles regardless of the system size.
   The system with equal temperature reservoirs is also a dissipative dynamical system due to these reservoir collisions.
   
  In the calculations reported here we use a system width $L_{y}=1.5$, a density $\rho =0.6$ and a fixed value of the right-hand  reservoir temperature $T_{R}=2$ and then choose the left-hand reservoir temperature $T_{L} > 2$ to obtain the required temperature gradient of $\nabla T=(T_{R}-T_{L})/L_{x}$ so there is a heat flow from left to right. 

   As this microscopic model couples mechanically and deterministically to the system it can be studied as a  dynamical system and also as a thermodynamic system with energy and kinetic entropy flows calculated numerically without approximations.
   These idealized heat reservoirs have been studied by computer simulation and kinetic theory \cite{KM09}, to obtain both heat conduction in low dimensional systems \cite{MT12,MT13}, and the Lyapunov spectrum and mode structure \cite{M12,MT13r}. 
   
   The instantaneous local temperatures for each particle are $T_{i,x}=p_{i,x}^{2}/m$ and  $T_{i,y}=p_{i,y}^{2}/m$ so the instantaneous system temperature is
\begin{equation} \label{kintemp}
T = \frac {1} {2N} \sum_{i=1}^{N} (T_{i,x}+T_{i,y})= \frac {1} {N} \sum_{i=1}^{N} \frac {\mathbf {p}_{i}^{2}} {2m},
\end{equation}	 
\noindent
   In the absence of a temperature gradient the average $\left< T \right>$ gives the system temperature, but when there is a temperature gradient the local time averages $\left<T_{i,x} \right>$ and $\left< T_{i,y} \right>$  give the local temperatures which will be used to determine the temperature profile inside the system.
    The difference between the components of the local temperature can be used to give a measure of the deviation from local thermodynamic equilibrium. 

   The deterministic reservoir allows the calculation of the usual dynamical systems properties \cite{Ott} as well the thermodynamic properties \cite{Zwanzig}.
   As the system contains hard disks of diameter $\sigma$ (which we set equal to 1) in a narrow channel of width $L_{y} < 2 \sigma$ that does not allow the disks to interchange positions, see figure (\ref{QDS}), we can associate any property of particle $i$ with the same {\it local} property of the volume element $V_{i} = L_{y}(\left< x_{i+1}-x_{i-1} \right>)/2$, centred at the average position of particle $i$, that is $\left<x_{i}\right>$.  
   Then for example, the local density is the inverse of the average volume occupied by the particle $\rho_{i} = 1/V_{i}$.
   This simple connection between volume element and particle property makes the QOD system optimal for these studies.
   
\subsection {Local Equilibrium}

   Where there is the possibility of a difference between the local values of the $x$ and $y$ temperatures, $T_{x}$ and $T_{y}$, we can modify the local-equilibrium distribution function as follows;
\begin{equation}\label{loc}
f_{loc} (x,{\bf v}) = \frac {m n} {2 \pi \sqrt{T_x T_y}}  \exp{\left[- \frac{m}{2} \left( \frac{v_x^2 } {T_x} + \frac{v_y^2 }{T_y}\right )\right]}.
\end{equation}
   Here $n$ is the local number density, and  $T_x$ and $T_y$ are the $x$ and $y$ components of temperature which are all functions of position $x$.  
   The kinetic entropy density obtained at the level of the local-equilibrium approximation is
\begin{eqnarray}\label{Bentdenx}
s_{loc}(x) &\simeq&  -\int d{\bf v} f_{loc}\ln f_{loc} \nonumber \\
&=& n \left[1 - \ln \left(\frac{m n}{2\pi}\right) + \frac {1} {2} \ln ( T_{x} T_{y}) \right]
\end{eqnarray}
which is reminiscent of the equilibrium Sackur-Tetrode equation except that the system is in a nonequilibrium steady state and the hydrodynamic fields are local.    

\subsection{Microscopic Heat flux vector}

   For a system of spherical particles the microscopic representation for the instantaneous local heat flux vector at position ${\bf r}$ and at time $t$ is given in \cite{EM2,McL89}.
   We restrict ourselves to the case where the local streaming velocity ${\bf u}({\bf r})$ is zero everywhere.  
   Defining the vectors $\mathbf {r}_{ij} =  \mathbf {r}_{j} -  \mathbf {r}_{i}$ and $\mathbf {p}_{ij} =  \mathbf {p}_{j} -  \mathbf {p}_{i}$, the impulse force at collisions between particles $i$ and $j$ is ${\bf F}_{ij} = ({\bf \hat r}_{ij} \cdot {\bf v}_{ij}) {\bf \hat r}_{ij} \delta (t- t_{ij})$, where $t_{ij}$ is the collision time and ${\bf \hat r}_{ij}$ is the unit vector in the direction of ${\bf r}_{ij}$.
   
   The total heat flux ${\bf J}_{Q} (t)$ is obtained as the volume integral of the local heat current ${\bf J}_{Q} ({\bf r}_{i},t)$ over the volume assigned to particle $i$, that is $V_{i}$.
   For the QOD system the local heat flux has potential contributions from two sources, either from a collision of particles $i$ and $i+1$ or from a collision of particles $i-1$ and $i$.
   The result is
\begin{eqnarray}\label {jqL}
&&{\bf J}_{Q} ({\bf r}_{i},t) V_{i} = U_{i} {\bf v}_{i}   \nonumber \\
&&- \frac {1}{4} \sum_{j \in \{i-1,i+1\}}^{N} {\bf \hat r}_{ij} ({\bf \hat r}_{ij} \cdot {\bf v}_{ij}) {\bf \hat r}_{ij}\cdot ({\bf v}_{i}+{\bf v}_{j})\delta (t- t_{ij}) 
\end{eqnarray}
where $U_{i} = \frac {1} {2} m({\bf v}_{i}- {\bf u}({\bf r}))^{2}$ is the internal energy of particle $i$.   
   In this form it is clear that the kinetic contribution is at ${\bf r}_{i}$ while there are two potential contributions, one at ${\bf r}_{i}$ and the other at ${\bf r}_{j}$.

   The time averages of the heat current must satisfy the continuity equation so on average, the same amount of heat passes through any vertical line regardless of its $x$ position.
   The heat current density in equation (\ref{jqL}) defines the instantaneous heat current at some arbitrary $x$ and $t$ but there is only a kinetic contribution if there is a particle at $x_{i} = x$, and there is only a potential contribution if two particles collide where one is at $x_{i} < x$ and for the other at $x_{j} > x$, so it is the time average of this instantaneous quantity that satisfies the continuity equation.


\section{Nonequilibrium}
\label{theory}

   In the same way as Green proposed that the equilibrium expression for the entropy could serve as reasonable approximation for the entropy in nonequilibrium systems, we will generalize equation (\ref{greenent}) to apply to nonequilibrium QOD systems with a steady heat flow.
   We consider a steady state system with a temperature gradient along the $x$ direction where the average total momentum is zero.
   Mechanical equilibrium implies that the pressure is constant throughout the system, but the imposed temperature gradient induces a density gradient and this needs to be explicitly included in the nonequilibrium form for the entropy.
   
   The temperature gradient implies a density gradient so $\rho$ is position dependent, and the distribution function $g_{2}$ for particle $i$, defined using $\rho ({\bf r})$, will ensure that the integrand goes to zero at large separations.
   Thus we propose the following expression for the nonequilibrium entropy of particle $i$
\begin{eqnarray} \label{2}
s^{\phi}_{i} &=& -\int f_{i} (v) \ln f_{i} (v) dv - \ln \rho({\bf r}_{i})   \nonumber \\
 &-& \frac {1} {2} \int d{\bf r}  \rho({\bf r}) (g_{2} ({\bf r}) \ln g_{2} ({\bf r}) - g_{2}  ({\bf r})+ 1) \nonumber \\
 \end{eqnarray}
where ${\bf r}=(r,y)$ is measured from the position of particle $i$.
   The integral is over the volume of the QOD system, $-L_{y}/2 < y < L_{y}/2$ and $-R_{c} < r < R_{c}$ where $R_{c}$ is a cut-off distance.
   Key to this form is that the density is position dependent and that the correlation functions $g_{2}$ are defined  so that $g_{2} \rightarrow 1$ as $r \rightarrow \infty$, so that $g_{2} \ln g_{2} - g_{2} + 1$ goes to zero ensuring the convergence of the integral.
   The first two terms in Eq. (\ref{2}) are the kinetic contributions to the entropy $s^{K}_{i}$ of particle $i$ and the subsequent terms are the configurational contributions $s^{\phi}_{i}$.
   
   The entropy density in the volume element associated with particle $i$ is then given by
\begin{eqnarray} \label{dens2}
s({\bf r}_i) &=& -\rho({\bf r}_{i}) \int f_{i} (v) \ln f_{i} (v) dv  - \rho({\bf r}_{i}) \ln \rho({\bf r}_{i})  \nonumber \\
 &-& \dfrac {1} {2} \rho({\bf r}_{i}) \int d{\bf r}  \rho({\bf r}) (g_{2} ({\bf r}) \ln g_{2} ({\bf r}) - g_{2}  ({\bf r})+ 1) \nonumber \\
 \end{eqnarray}
  This nonequilibrium entropy density varies across the QOD system as the velocity distribution changes, the local density $\rho({\bf r}_{i})$ changes and the correlation function $g_{2} ({\bf r})$ also changes from particle to particle.
  In particular, the nonequilibrium boundary conditions induce an asymmetry in $g(r,y)$ with respect to a sign change in $r$.
   We will look more closely at these contribution in the coming sections.  
     
\subsection {Configurational entropy}

   We study nonequilibrium QOD systems of $80$, $160$ and $320$ hard disks at a density of $\rho=0.6$ with system width $L_{y}=1.5$.
   The temperature of the cold reservoir on the right-hand side is kept constant at  $T_{R}=2$, and different values of the temperature of the left-hand reservoir $T_L$ are used to obtained the required temperature gradient $\nabla T$.
  For $N=80$ we consider three values for the temperature of the hot reservoir, $T_{L}=2$ so $\nabla T=0$, $T_{L}=18$ a moderate gradient $\nabla T=-0.18$ and $T_{L}=34$ a high gradient $\nabla T=-0.36$.
  For each of the larger system sizes we consider the same values of $\nabla T$.

   It has previously been observed that the numerically calculated local {\it kinetic} entropy density agrees well with that calculated from the local equilibrium approximation \cite{MT13}. 
    While the momentum distributions for the particles cannot be exactly Gaussian for a nonequilibrium steady state, the deviations from Gaussian are at best only subtle and the local kinetic entropy density calculated from the numerical distributions is almost indistinguishable from the local kinetic entropy density obtained from the local equilibrium distribution.

\begin{figure}[htb]
\begin{center}
\caption{(color online) The full two-dimensional distribution function $g(r,y)$ for particle $20$ in a system of $80$ disks at a density of $0.6$. The color scale is shown on the right-hand side. The hot reservoir temperature is $T_{L}= 18$ so $\nabla T = -0.18$. The local temperature at particle $20$ is $8.62$ and the local density is $0.557$.  }	\label{gry20}
	\includegraphics [ scale=0.6] {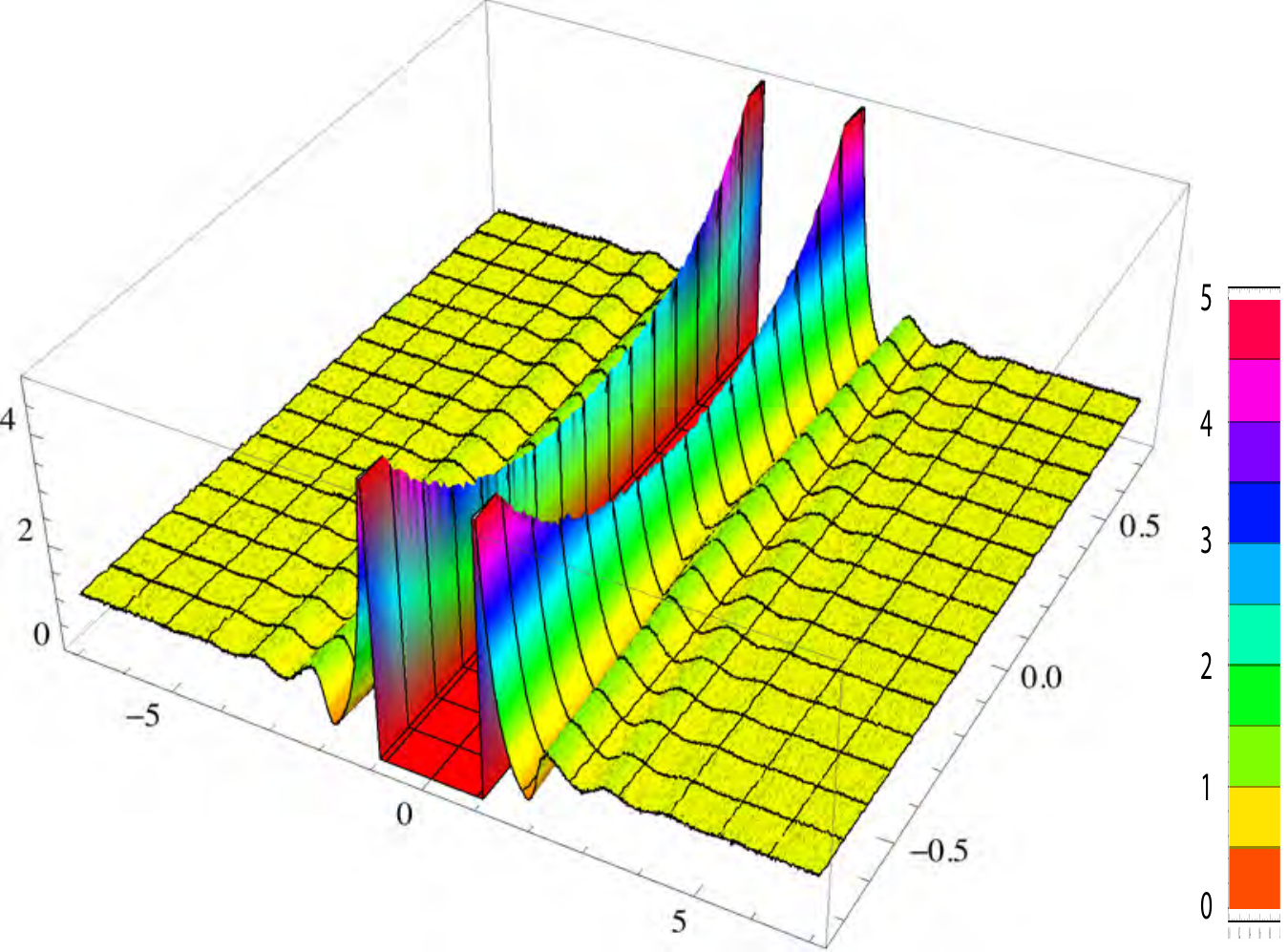}
\end{center}
\end{figure}

\begin{figure}[htb]
\begin{center}
\caption{(color online) The full two-dimensional distribution function $g(r,y)$ for particle $60$ in a system of $80$ disks at a density of $0.6$. The color scale is the same as Fig. (\ref{gry20}). The hot reservoir temperature is $T_{L}= 18$ so $\nabla T = -0.18$. The local temperature at particle $60$ is $4.49$ and the local density is $0.664$.  }	\label{gry60}
	\includegraphics [scale=0.6] {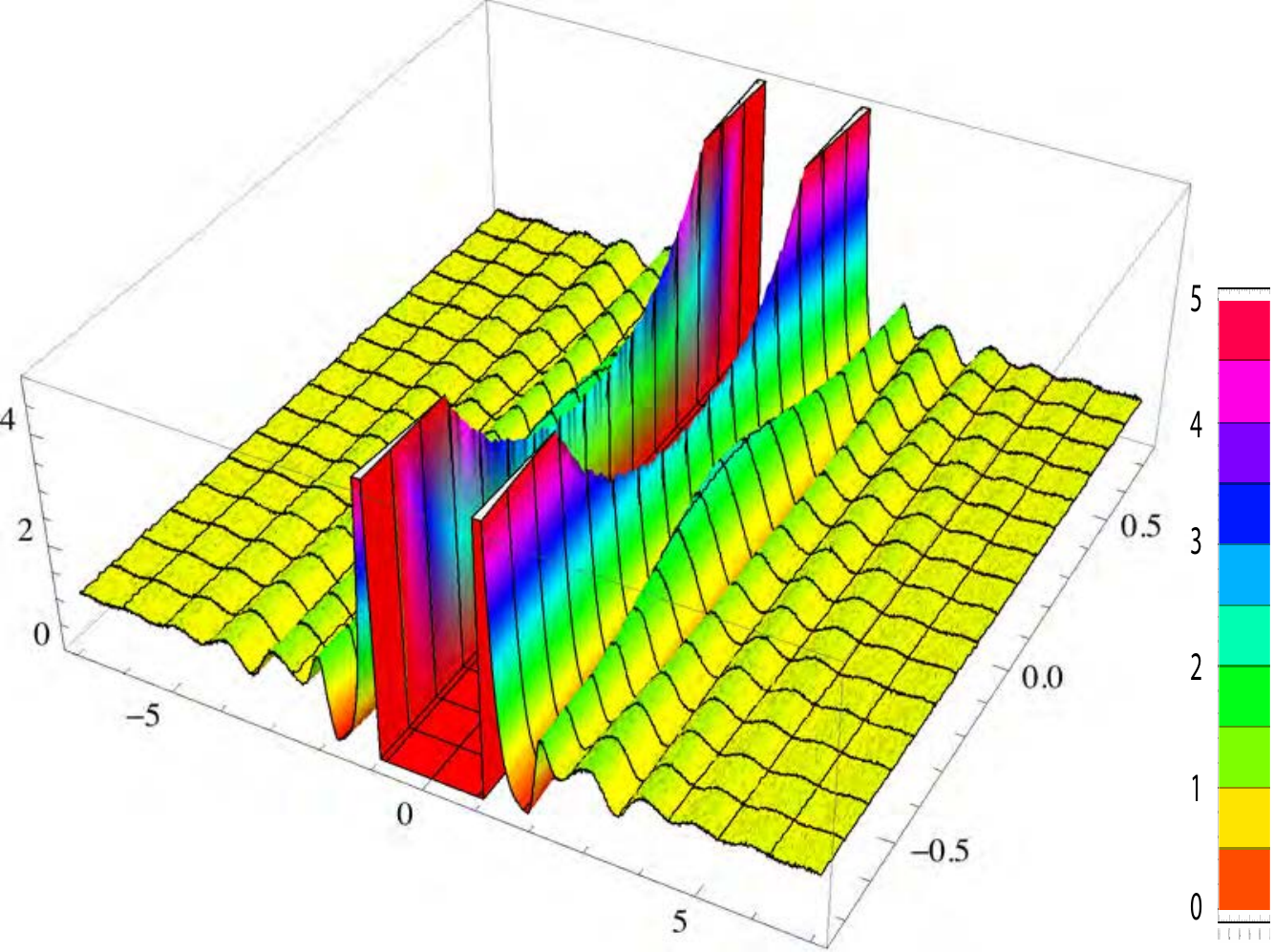}
\end{center}
\end{figure}

   The configurational contribution is calculated from Eq. (\ref{2}) using the two-particle correlation function $g(r,y)$ collected on a histogram of $1400 \times 150$ bins where $r$ is the radial distance $\sqrt {x^2 + y^2}$.
   This corresponds to a cutoff in the integral over $r$ of $R_{c} = 7$.
   For a QOD system the correlation function can be written as a function of $r$ and $y$ so the $2$-particle configurational integral becomes
\begin{eqnarray} \label{gry}
s^{\phi}_{i} &=&- \frac {1} {2} \int_{-R_{c}}^{R_{c}} dr  \int_{-L_{y}/2}^{L_{y}/2} dy \big (\frac {\partial x} {\partial r}\big)  \rho (g \ln g - g + 1) \nonumber \\
\end{eqnarray}
where $\rho = \rho (r_{i})$ and $g=g_{i}(r,y)$.
   The slope of the density gradient is estimated locally at the position of each particle and a linear approximation to the density gradient is used in the integral. 
   This can lead to inaccuracies near the boundaries where the density varies quickly and possibly nonlinearly on $r$.
   A typical result for $g(r,y)$ for particle $20$ in an $80$ particle system is shown in Fig. (\ref{gry20}) with some strong $y$ dependence at contact ($r=1$) but this disappears quickly at larger $r$.
   For particle $60$ in the same system $g(r,y)$ is shown in Fig. (\ref{gry60}) and the configurational entropy from $g(r,y)$ is larger in magnitude and the second neighbour peak varies more strongly with $y$. 

   The integral in Eqn. (\ref{gry}) has a maximum cutoff of $R_{s}=7$ particle diameters and the convergence of the result depends on the smoothness of the numerically generated $g(r,y)$.
   Again near the reservoir boundaries the cutoff needs to be limited so that the range of the integral remains within the simulation cell.
   If $g(r,y)=g_0 (r,y) + \Delta(r,y)$ where $g_0 (r,y)$ is the exact result and $\Delta(r,y)$ is a noise term that goes to zero $\propto N_{s}^{-1/2}$ where $N_{s}$ is the number of samplings of the correlation function and the integral of $\Delta(r,y)$ over $r$ and $y$ is zero, then it can be shown that $s = s_{0} - O(N_{s}^{-1})$. 
   This implies that $s$ is a lower bound on the exact value $s_{0}$.
   
   In Table (\ref{Int_converg}) we present a calculations of the configurational entropy as a function of the number of histogram samplings $N_{s}$ and the integral cut-off $R_{c}$ used in Eq. (\ref{gry}).
   In each case the differences are quite small with the maximum being $0.25\%$. 
   
\begin{table}[htdp]
\caption{The total configurational entropy $s^{\phi}_{i}$ of particle $i$ in a QOD system of $80$ disks at a density $\rho = 0.6$ with the right-hand  reservoir temperature of $T_{R}=2$ and a left-hand reservoir temperature $T_{L}=18$ with a correlation function cut-off of $R_{c} = 7$. The column headed $N_{s}=2000$ is the converged result, the column headed $N_{s}=1000$ is the result with only half the sampling and the last column is the converged result with $R_{c} = 6$ (rather than $7$). The largest percentage difference is 0.25 \%. }
\begin{center}
\begin{tabular}{|c|c|c|c|} \hline
particle \#    &  $N_{s}=2000$   &  $N_{s}=1000$    &  $R_{c} = 6$   \\  \hline
$20$             &  $ -1.02264    $   &  $-1.02398     $  &  -1.02005          \\
$40$             &  $ -1.28851    $   &  $-1.29086     $  &  -1.28569          \\
$60$             &  $ -1.72306    $   &  $-1.72550     $  &  -1.71994          \\
\hline
\end{tabular}
\end{center}
\label{Int_converg}
\end{table}%

\subsection{Radial correlations}

   If the full correlation function $g(r,y)$ only depends strongly on $y$ at contact we can calculate an approximate configuration entropy using a reduced correlation function $g(r)$ where the $y$ dependence of $g(r,y)$ has been integrated.
   The entropy density calculated using $g(r,y)$ and $g(r)$ agree where the dependence of $g$ on $y$ is small and the local density is smallest, but when the density increases the local packing of hard disks  begins to dominate the structure and the result from $g(r,y)$ is the most accurate.
   As we can distinguish positive and negative values of $x$ the reduced distribution $g(r)$ is defined for $-R_{c}<r<R_{c}$ and the configurational entropy contribution is
\begin{eqnarray} \label{grr}
s^{\phi,r}_{i} &=&- \frac {L_{y}} {2} \int_{-R_{c}}^{R_{c}} dr   \rho (g \ln g - g + 1) 
\end{eqnarray}
      This correlation function has much better statistics than the full $g(r,y)$ and this allows us to investigate the asymmetry induced in $g(r)$ by the nonequilibrium boundary conditions.

\begin{figure}[htb]
\begin{center}
\caption{(color online) The local configurational entropy $s^{\phi}_{i}$ calculated from the distribution function $g(r,y)$ (red symbols) and calculated from the integrated distribution $g(r)$ (blue symbols) plotted as a function of the particle number for systems of  $80$ disks at a density of $0.6$. The label on each curve is the left-hand temperature $T_L$. }	\label{s_den}
	\includegraphics{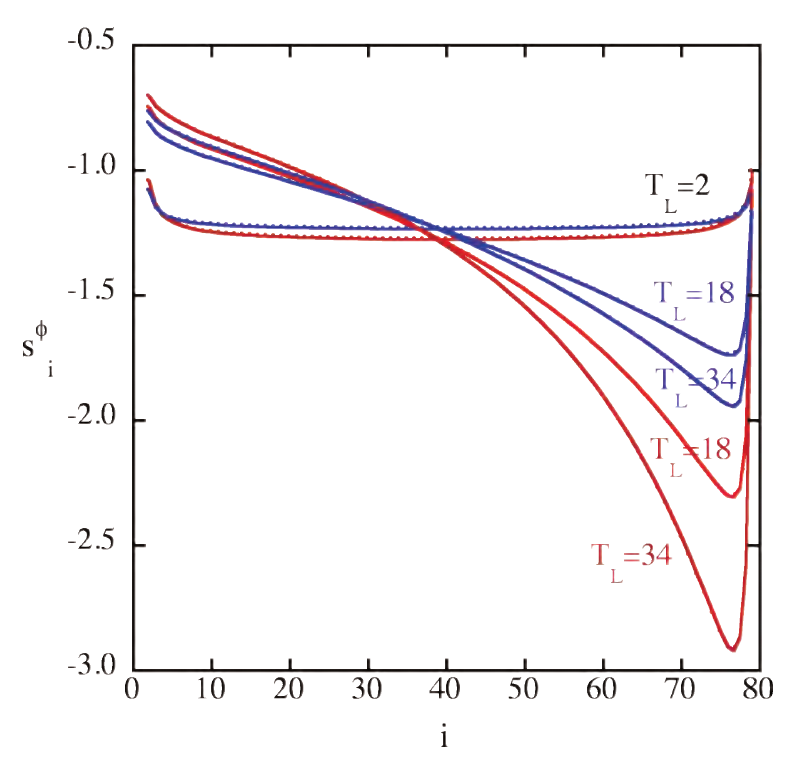}
\end{center}
\end{figure}

   The results in Fig. (\ref{s_den}) show the configurational entropy density $s^{\phi}_{i}$ for systems of $80$ disks for three different values of $\nabla T$.
   The results obtained from $g(r,y)$ differ from those from $g(r)$ particularly near the cold reservoir on the right-hand side.
   Here the density is highest and hard disk packing effects lead to a stronger dependence on $y$.
   As the value of $\nabla T$ increases the total configurational entropy decreases despite the fact that there are both increases and decreases in the entropy locally.
   
   A temperature gradient breaks inversion symmetry in the $x$ direction so the integral can be split into a positive $r$ region and a negative $r$ region where the integrands are different.
   If $g^{+}(r)=g(r)$ and $g^{-}(r)=g(-|r|)$ for $r>0$, we can define new functions $g^{m}=\sqrt {g^{+} g^{-}}$ the multiplicative mean and $g^{r}=\sqrt {g^{+}/g^{-}}$ so that the integrand in Eqn. (\ref{gry}) becomes
\begin{eqnarray}
&&g^{+} \ln g^{+} -g^{+} +1 + g^{-} \ln g^{-} -g^{1} +1  \nonumber \\
&=& (g^{r} + 1/g^{r})\{g^{m} \ln g^{m} -g^{m} +1\}   \nonumber \\
&+& (g^{r} - 1/g^{r}) g^{m} \ln g^{r} + 2 - (g^{r} + 1/g^{r})
\end{eqnarray}   
and the integral is now over $r$ positive. 
  The first term is an {\it equilibrium} like contribution and the second term is a measure of the asymmetric contribution. 
  Defining a function $\Delta g(r)$ by $g^{+} (r)=g^{-}(r) (1 + \Delta g(r))$ it can be shown that 
\begin{eqnarray} \label{g^r}
g^{r} + 1/g^{r} &=& 2 + \frac {\Delta g^{2}} {4} + O(\Delta g^{3}),  \nonumber \\
g^{r} - 1/g^{r} &=& \Delta g - \frac {\Delta g^{2}} {2}+ O(\Delta g^{3})
\end{eqnarray}   
so the integrand becomes
\begin{equation} \label{intgm}
2 \{g^{m} \ln g^{m} -g^{m} +1\} +  \{g^{m} \ln g^{r} \} \Delta g + O(\Delta g^2)
\end{equation}   
   The leading term is of the same form as the equilibrium term, but with $g^{m}$ replacing $g$. 
   This is just one of many possible symmetric functions that can be constructed from $g^{+}$ and $g^{-}$. 
   Here we are interested in an estimate of the size of the  asymmetric contribution to the entropy although it is still possible that the numerical difference between $g^{m}$ and $g$ may also contribute.
     
\begin{figure}[htb]
\begin{center}
\caption{(color online) The correlation function $\Delta g_{120} (r)$ for particle $120$ in $160$ disk QOD systems at a density $\rho = 0.6$, a right-hand temperature of $T_{R} = 2$, for different values of the left-hand temperature $T_{L}$. The red line is for $T_{L}=2$ so no temperature gradient, the blue line is for $T_{L}=34$ and the green line is for $T_{L}=66$.}	\label{del_160}
	\includegraphics{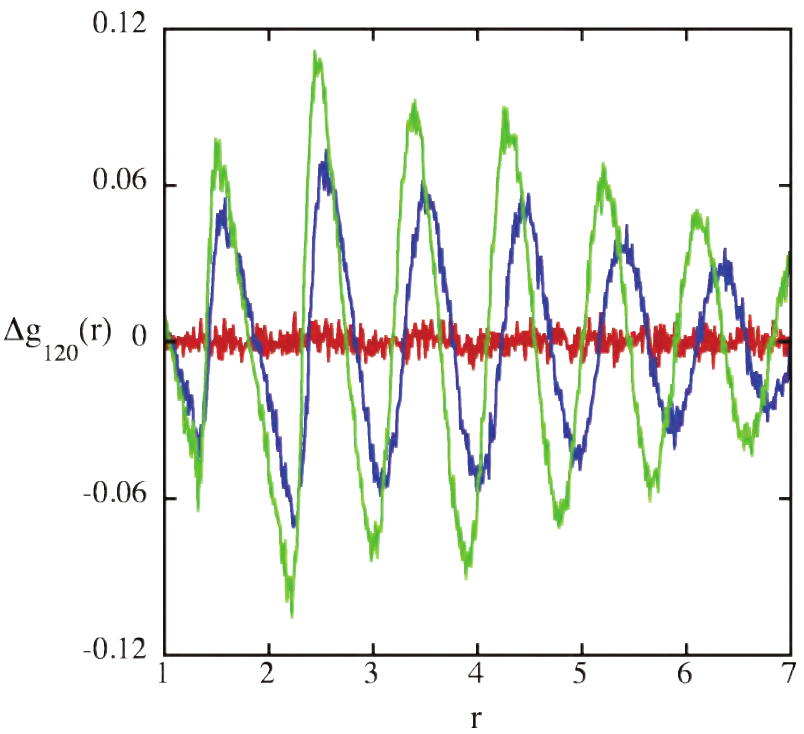}
\end{center}
\end{figure}

\begin{figure}[htb]
\begin{center}
\caption{(color online) The correlation function $\Delta g_{150}$ for particle $150$ in $160$ disk QOD systems at a density $\rho = 0.6$, a right-hand temperature of $T_{R} = 2$, for different values of the left-hand temperature $T_{L}$. The red line is for $T_{L}=2$ so no temperature gradient, the blue line is for $T_{L}=34$ and the green line is for $T_{L}=66$.}	\label{dgm_160}
	\includegraphics{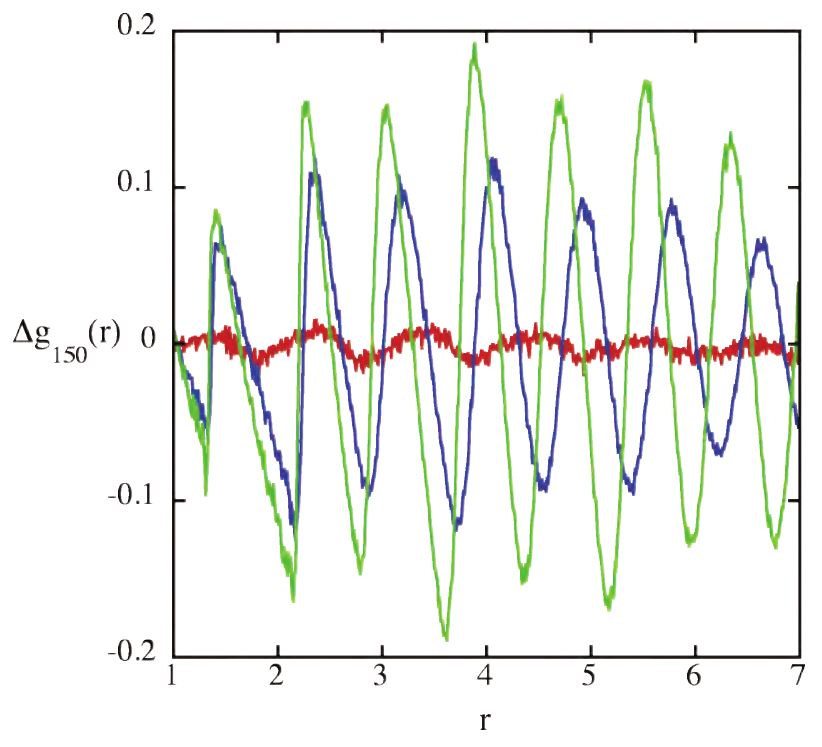}
\end{center}
\end{figure}

   The asymmetry imposed on the system by the nonequilibrium boundary conditions is easily seen in Fig. (\ref{del_160}) where we consider $\Delta g(r)$ for particle $120$ in a $160$ particle QOD system.
   This particle is towards the cold reservoir where the configurational entropy is greatest so we expect a significant signal which increases with temperature gradient.
   Here we see a strong contribution which is very long-ranged extending well past the sampling window limit at $r=7$ despite this the integral result we obtain for $\Delta s_{r}$ seems well converged.
   The graph of $\Delta g(r)$ for particle $150$ in Fig. (\ref{dgm_160}) shows similar behaviour to that observed in Fig. (\ref{del_160}) except that here the vertical scale is larger and we see oscillations in $\Delta g(r)$ at $\nabla T = 0$ due to the asymmetry imposed by the nearness of the hard wall boundary.
   The function $\Delta g(r)$ gives a good measure of the asymmetric contribution from either the nonequilibrium boundary conditions or the closeness to a boundary.


   The integrand obtained in Eq. (\ref{intgm}) has to be multiplied by $\rho (r)$ before integrating and here we use a local linear approximation of the form $\rho (r) = \rho_{0} + \Delta \rho r$.
   This means that the first contribution in Eq. (\ref{intgm}) multiplied by $\rho_{0}$.
   The next contribution comes from the slope  $\Delta \rho$. 
   Here again the dominant term is determined by the sizes of the prefactors $ \{ g^{r} -1 / g^{r} \}$ and $ \{ g^{r} + 1 / g^{r} \}$ which from Eq. (\ref{g^r}) are $\Delta g$ and $2 + O(\Delta g^2)$ respectively.   
   The final result to leading order is
\begin{eqnarray} \label{entdel}
s^{\phi,m}_{i}&=&-L_{y}  \rho_{0} \int_{0}^{R}dr  (g^{m} \ln g^{m} - g^{m} + 1).  \nonumber \\
\end{eqnarray}   
   with first order corrections that are linear in $\Delta \rho$ and $\Delta g(r)$. 
   This is essentially the original expression for the configurational entropy Eq. (\ref{gry}) with $g^{m}$ replacing $g$.
   Only the first term on the first line has a term that does not have a contribution from the difference between $g^{+}$ and $g^{-}$ or from the slope of the density profile.

   The expression for the nonequilibrium entropy given in Eq. (\ref{2}) contains contributions from changes in velocity distributions, density profiles and the asymmetry of $g(r)$.
   Although we can separate the kinetic changes due to changes in the velocity distributions, we have been unable to completely separate the effects of density profiles from asymmetries in $g(r)$ expect that it appears that asymmetry has a very small effect.
   In Table \ref{asym} we estimate the contribution to the local entropy from the asymmetry of the distribution function $g(r)$ and the slope of the density $\Delta \rho$ by calculating the difference $s^{\phi,r}_{i} - s^{\phi,m}_{i}$.
   While in general the differences are very small they can be as large as $1-2\%$ near the cold reservoir.
\begin{table}[htdp]
\caption{Estimating the anisotropy contribution to the local configurational entropy $s^{\phi,r}_{i} - s^{\phi,m}_{i}$ for a QOD system of $80$ disks at a density $0.6$ at different values of $\nabla T$ and different particle numbers.}
\begin{center}
\begin{tabular}{|c|ccc|} \hline
$\nabla T$ & $20$ &   $40$   & $60$  \\  \hline
$-0.18$      & $-0.00007$&   $-0.00018$   & $-0.00069$    \\
$-0.36$      & $-0.00008$&   $-0.00027$   & $-0.00145$    \\
\hline
\end{tabular} \end{center} \label{asym}
\end{table}

\subsection{System size scaling}

   Our purpose here is to identify approximate (or possibly exact) scaling relations for the properties of the system a function of system size $N$ and temperature gradient $\nabla T$, with a view to separating bulk properties from surface properties (or boundary effects).
   To look more closely at system size scaling we consider three systems with the same temperature gradient $\nabla T = -0.18$ with $80$, $160$ and $320$ disks.
   As the right-hand temperature is fixed at $T_{R}=2$, we can take the temperature profile for $80$ particles and multiply the $T_x$ axis by two and the $x$ axis by $1.525$ to match the $160$ particle profile.
   Similarly, taking the $160$ particle profile, multiplying the $T_x$ axis by two and the $x$ axis by 1.666, matches the $320$ particle $T$ profile.
   
\begin{figure}[htb]
\begin{center}
\caption{(color online) The scaling of the density profiles for $80$ (red symbols) and $160$ (blue symbols) QOD systems particles onto the profile for $320$ (green symbols) particles at a density $\rho = 0.6$. In each case the temperature profile is $\nabla T = -0.18$. }	\label{T_sc}
	\includegraphics{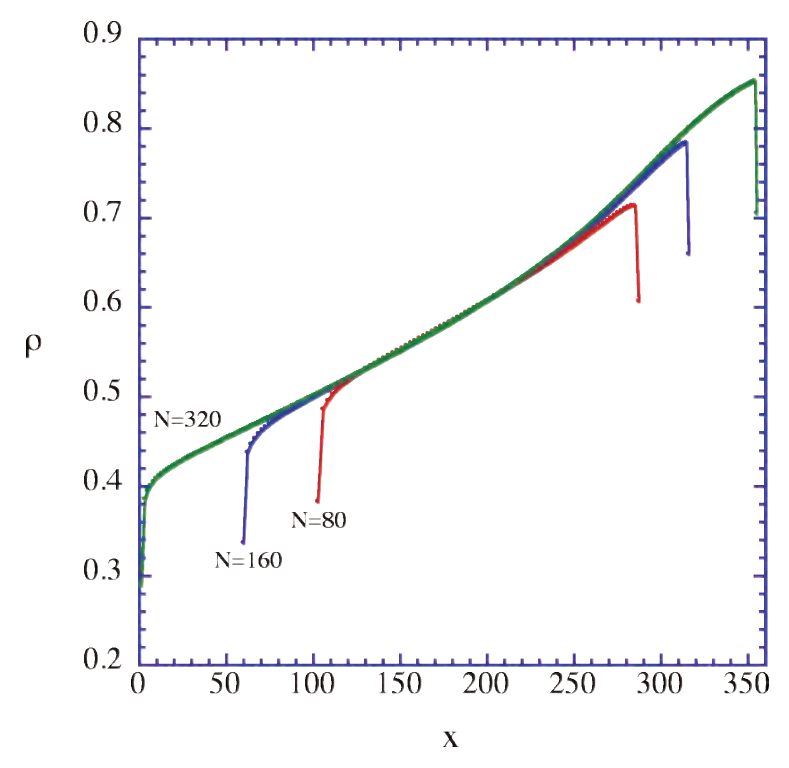}
\end{center}
\end{figure}

   The average density of the system is $0.6$ so this sets the mean of the density profile which occurs in a volume element close to the centre of the system.
   Scaling the $x$ axis for the $80$ particle system using $x_{160} = 16\times (x_{80} +20)/11$ maps the $80$ particle profile onto the $160$ particle profile.
   Again, using the scaling  $x_{320} = 16\times (x_{160} +40)/11$ maps the $160$ particle profile onto the $320$ particle profile.
   The results for this scaling are shown in Fig. (\ref{T_sc}).

\begin{figure}[htb]
\begin{center}
\caption{(color online) The scaling of the configurational entropy profiles for $80$ and $160$ QOD systems particles onto the profile for $320$ particles at a density $\rho = 0.6$ and a temperature profile of $\nabla T = -0.18$. The symbols are the same as Fig. (\ref{T_sc}), red for $80$, blue for $160$ and green for $320$.}	\label{DS_sc}
	\includegraphics{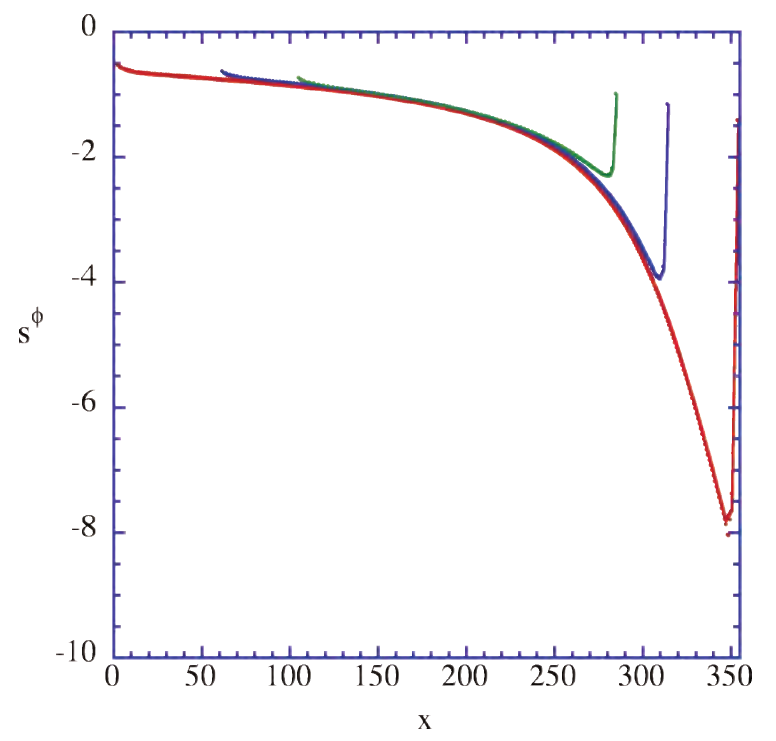}
\end{center}
\end{figure}

   The configurational entropy of each particle can also be scaled in the same way as the density.
   The $x_{80}$ coordinate for $N=80$ is scaled by $x_{160} = 16\times (x_{80} +20)/11$ and then by $x_{320} = 16\times (x_{160} +40)/11$, while for the $N=160$ system we use only the second of these scalings, to map these results onto those for $N=320$.
   Essentially the deviations are only noticeable near the system boundaries.
   The results shown in Fig. (\ref{DS_sc}) are as good as those for the density profile in Fig. (\ref{T_sc}).
   This means the volume elements with the same density have the same configurational entropy density regardless of system size and increasing system size adds regions to the left and right hand sides that have larger deviations from the mean density and the central value of configurational entropy.
   The $K$-particle correlation functions for a system of hard core particles only depends on the density and not on the temperature so the configurational contribution to the entropy scales with density.
   By contrast, the local equilibrium kinetic entropy density which is observed to be almost equal to the calculated kinetic entropy density, scales with both the temperature and density, thus the two contributions to the total entropy density scale differently.  

\subsection{Total Entropy Density}
   
\begin{figure}[htb]
\begin{center}
\caption{(color online) The components of the total entropy density for an $80$ particle QOD system with $\nabla T =-0.36$. The green symbols are the calculated kinetic entropy density and the green crosses are the local equilibrium result. These two results are indistinguishable on this scale. The red symbols are the configurational contribution to the entropy density and the blue symbols are the total entropy density.}	\label{s(x)}
	\includegraphics{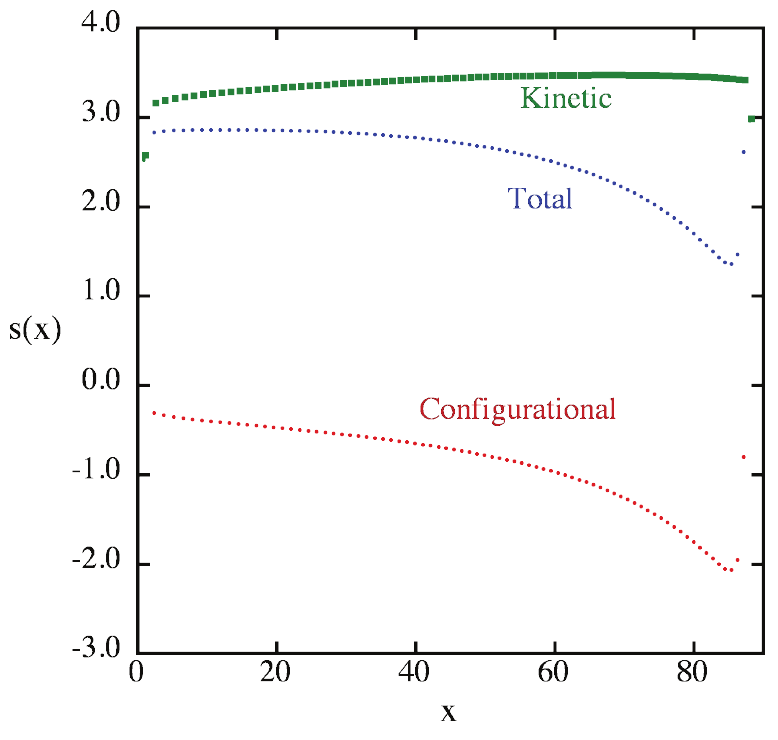}
\end{center}
\end{figure}
   
   Combining the previously calculated kinetic entropy densities with the configurational entropy density calculated here we can consider the relative sizes of the components.
   In Fig. (\ref{s(x)}) the results for an $80$ particle system at a density $0.6$ and system width $1.5$ with $\nabla T = -0.36$ are shown.
   The kinetic entropy density is always positive and the configurational entropy density is always negative but smaller in magnitude so the total entropy density remains positive.

\section{Conclusion}

   In this paper we have followed the prescription originally imagined by Green \cite{Green}, (i) that Eq. (\ref{2}) is correct when the fluid is in thermal and mechanical equilibrium, and (ii) when the fluid is not in thermal and mechanical equilibrium the entropy is always less than the equilibrium value.
   Strictly, Green was speaking about the canonical equivalent to Eq. (\ref{2}) rather than the grand canonical result used here.
   The demonstration of point (i) has been made previously in the work of Baranyai and Evans \cite{BE91} where both 2 and 3-body correlation functions are considered for equilibrium systems.
   Here we demonstrate point (ii) for nonequilibrium systems and more strongly by calculating the local entropy per particle or alternatively by calculating the local entropy density.
   The QOD system is particularly useful in connecting the property of a single particle with the density of that property in the average volume element occupied by that particle.
   
   It is interesting to remark that the derivations which lead from the entropy in Eq. (\ref{1}) to entropy per particle in Eq. (\ref{2}) rely on the canonical and grand canonical ensemble formulations that only apply at equilibrium.
   There is no equivalent derivation for a nonequilibrium steady state.
   Applying Eq. (\ref{2}) outside of equilibrium is based on the fact that the quantities involved, the local density $\rho$ and the 2-body correlation function $g_{2} ({\bf r})$ and higher order correlation functions, are well defined structural quantities that are clearly calculable in a molecular dynamics computer experiment.
   
   As Green has cautioned, the quantity we have calculated as the local nonequilibrium entropy has the properties that we expect, it is equal to the entropy at equilibrium, but it may not be the thermodynamically meaningful nonequilibrium entropy.
   It is an order parameter that is proportional to the nonequilibrium entropy and its usefulness in a thermodynamic context is yet to be determined.
   
\section*{Acknowledgements}
The author thanks Thomas Dean, Stuart Hatzioannou and David Webb for discussions and assistance with computations.

\bibliographystyle{entropy}
\makeatletter
\renewcommand\@biblabel[1]{#1. }
\makeatother
\bibliography{}

\end{document}